\documentclass[11pt]{article}
\usepackage{paclic29}
\usepackage{times}
\usepackage{latexsym}
\usepackage{amsmath}
\usepackage{multirow}
\usepackage{url}
\usepackage{graphicx}
\usepackage{placeins}
\graphicspath{ {Images/} }

\title{Understanding Rating Behaviour and Predicting Ratings by Identifying Representative Users}

\author{Rahul Kamath \\
  University of Tokyo \\
  7-3-1 Hongo, Bunkyo-ku \\
  Tokyo, Japan \\ \\\And
  Masanao Ochi \\
  University of Tokyo \\
  7-3-1 Hongo, Bunkyo-ku \\
  Tokyo, Japan \\ \\\And
  Yutaka Matsuo \\
  University of Tokyo \\
  7-3-1 Hongo, Bunkyo-ku \\
  Tokyo, Japan \\ \\}

\date{}

\begin{document}
\maketitle
\begin{abstract}
Online user reviews describing various products and services are now abundant on the web. While the information conveyed through review texts and ratings is easily comprehensible, there is a wealth of hidden information in them that is not immediately obvious. In this study, we unlock this hidden value behind user reviews to understand the various dimensions along which users rate products. We learn a set of users that represent each of these dimensions and use their ratings to predict product ratings. Specifically, we work with restaurant reviews to identify users whose ratings are influenced by dimensions like `Service', `Atmosphere' etc. in order to predict restaurant ratings and understand the variation in rating behaviour across different cuisines. While previous approaches to obtaining product ratings require either a large number of user ratings or a few review texts, we show that it is possible to predict ratings with few user ratings and no review text. Our experiments show that our approach outperforms other conventional methods by 16-27\% in terms of RMSE.
\end{abstract}

\section{Introduction}
\label{sec:intro}
With the advent of Web 2.0, a large number of platforms including e-commerce sites, discussion forums, blogs etc. have emerged that allow users to express their opinions regarding various businesses, products and services. These opinions are usually in the form of reviews, each consisting of text feedback describing various aspects of the product along with a single numeric rating representing the users' overall sentiment about the same \cite{mcauley2012learning}. Such user review ratings are normally aggregated to provide an overall product rating, which help other people form their own opinion and help them make an informed decision during purchase. However, in case of new products, there is a time delay till a sufficient number of ratings that give a `complete picture' of the product can be obtained. In such a scenario, the seller of the product may find it useful to identify a few people whose ratings, when combined together, reflect this `complete picture'. The seller may then invite these people to review the product and, as a result, reduce the time delay involved in getting the `true' product rating.

Review text is unstructured and inherently noisy. But it can be a valuable source of information since users justify their ratings through such text \cite{mcauley2013hidden}. Users tend to express their sentiments about different aspects of a product in the review text and provide a rating based on some combination of these sentiments \cite{ganu2009beyond}. However, some users are influenced heavily by one particular aspect of the product and this is reflected in their ratings. For example: While reviewing smartphones, the ratings provided by a user may be influenced heavily by just the battery-life, irrespective of the quality of other aspects of the phone. Similarly, while reviewing restaurants, some users' ratings may correlate with the ambience of the restaurant or the level of service provided. We call such users as `representative users' since their ratings tend to `represent' one particular dimension of the product. 

Although latent factors obtained from ratings data have been used extensively for rating prediction,

\FloatBarrier
\begin{figure*}[t]
\centering
\includegraphics[scale=0.11]{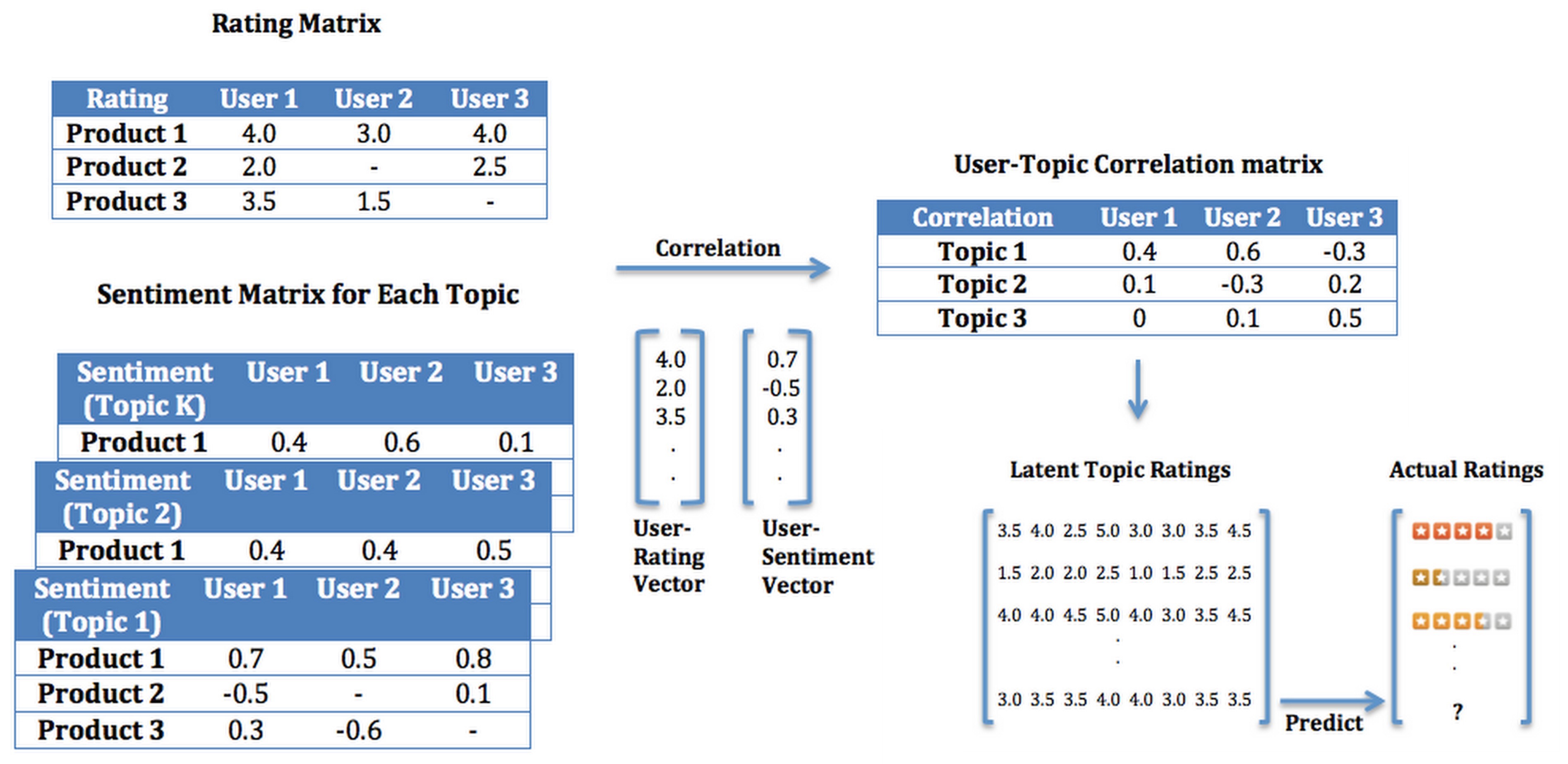}
\caption{An overview of our proposed method}
\label{figure:overview}
\end{figure*}
\FloatBarrier

\noindent very few previous works have attempted to combine both review text and ratings. Our approach combines latent topics obtained from review text with users' rating data to learn representative users for each product. This enables us to predict ratings for new products by just looking at the ratings of a small set of users, even when no review text is available. In traditional methods, product ratings are obtained by modelling the product factors from ratings data. However, \cite{mcauley2013hidden} suggest that this approach is not accurate in case of new products due to the lack of sufficient number of ratings. They, in turn, propose a model which fits product factors from a few review texts. Our approach is free from both these constraints.

In this study, we use the topic model Multi-Grain Latent Dirchlet Allocation (MG-LDA) described in \cite{titov2008modeling} on restaurant reviews obtained from Yelp\footnote{\url{http://www.yelp.com}} to obtain latent topics that correspond to ratable aspects of the restaurants. Since we segregate the reviews on the basis of restaurant category, we notice some interesting variations across different cuisines. The words associated with the extracted topics are then used to perform review segmentation where we identify the sentences that describe each topic. This also enables us to analyse the sentiment expressed regarding each topic in a review. We then capture the intuition of representative users to learn a set of users who best represent each topic. Latent topic ratings for restaurants are then obtained by aggregating the ratings of those users who represent that topic. The overall ratings of new restaurants are then predicted using a regression model. An overview of the proposed method is shown in Figure~\ref{figure:overview}.

We also show how this concept could be used to better understand rating behaviour across different cuisines. For example: What do people who visit French restaurants care most about - food, service or value for money? How is this different from people who visit Italian restaurants?

The rest of the paper is structured as follows. Section~\ref{sec:papers} provides a review of related work. Section~\ref{sec:methodology} describes our proposed method. In Section~\ref{sec:experiments}, we describe the experiments performed and report the results of our evaluation. Section~\ref{sec:end} concludes the paper with a summary of the work and the scope for future work.

\section{Related Work}
\label{sec:papers}

One of the earliest attempts at rating prediction that combines both review text and ratings is \cite{ganu2009beyond}. However, their review segmentation method differs from ours in that their work depends on manual annotation of each review sentence into pre-determined domain-specific aspects and the training of separate classifiers for each aspect. Furthermore,

\FloatBarrier
\begin{figure*}[t]
\centering
\includegraphics[scale=0.5]{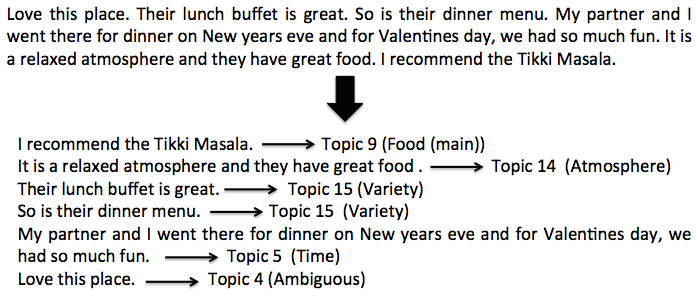}
\caption{Review Segmentation}
\label{figure:segmentation}
\end{figure*}
\FloatBarrier

\noindent  it does not capture the variation that may exist within the domain. For example: The aspects that affect ratings for French restaurants (e.g. `Drinks (wine)', `Deserts' etc.) may be different from those of Indian restaurants (e.g. `Flavour (spiciness)', `Variety' etc.). \cite{wang2010latent} approach the problem of segmentation by measuring the overlap between each sentence of the review and the seed words describing each aspect. However, these aspect seed words are chosen manually which are, again, domain-specific. 

Topic models are normally used to make the segmentation task transferable across different domains. The problem of mapping such topics into aspects is studied in \cite{titov2008joint,lu2011multi,brody2010unsupervised,mcauley2012learning,jo2011aspect}. \cite{titov2008joint,mcauley2012learning} use explicit aspect ratings as a form of weak supervision to identify rated aspects while \cite{lu2011multi} use manually selected aspect seed words as a form of weak supervision. To remove the dependence on aspect ratings and aspect seed words, \cite{jo2011aspect} develop a model that captures aspects using a set of sentiment seed words while \cite{brody2010unsupervised} present an unsupervised method for extracting aspects by automatically deriving the sentiment seed words from review text. It is important to note that we do not map the latent topics we obtain into explicit aspects since it is not necessary for our final goal. 

Rating prediction is also studied in \cite{gupta2010capturing,moghaddam2011ilda,baccianella2009multi} where the authors focus on multi-aspect rating prediction and in \cite{mcauley2013hidden} where the authors build a recommendation system using a combination of latent dimensions obtained from rating data and latent topics obtained from review text. 

\section{Methodology} 
\label{sec:methodology}

\subsection{Dataset and Preprocessing}
\label{ssec:data}

We use the Yelp Challenge Dataset\footnote{\url{http://www.yelp.com/dataset_challenge}} consisting of around 1.12 million reviews of more than 42000 restaurants across 4 countries. These reviews are provided by more than 250000 users. Reviews contain a single star rating, text, author etc. Details of restaurants like average star rating, categories (cuisine) etc. are also available.  We segment the restaurants according to its category since we would like to better understand the variation that exists across different cuisines. Note that we ignore the fact that certain restaurants may have multiple categories. For example: Some Indian restaurants may also serve Thai food.

We tokenize the review text along whitespaces, remove all punctuation and stop-words, and lemmatize the words using the NLTK Wordnet lemmatizer described in \cite{bird2009natural}.

\subsection{Topic Extraction}
\label{ssec:mglda}

We run the topic model multi-grain LDA described in \cite{titov2008modeling} on a corpus of restaurant reviews obtained from a single cuisine to extract $K$ latent topics. Unlike standard topic

\FloatBarrier
\begin{table*}[t]
\begin{tabular}{|c|c|l|}
\hline
Cuisine & Interpreted Topic & \multicolumn{1}{c|}{Top Words} \\ \hline
\multirow{5}{*}{Indian} & Variety & buffet,lunch,dish,vegetarian,menu,selection,option,good,item,great \\ \cline{2-3} 
 & Food & chicken,masala,tikka,curry,naan,lamb,dish,paneer,tandoori,ordered \\ \cline{2-3} 
 & Flavour & spicy,spice,flavour,dish,hot,curry,food,like,sauce,taste \\ \cline{2-3} 
 & Value & price,portion,food,meal,get,two,small,little,rice,bit \\ \cline{2-3} 
 & Atmosphere & restaurant,place,nice,decor,inside,strip,little,clean,like,table,look \\ \hline
\multirow{5}{*}{Italian} & Food (Pizza) & pizza,crust,good,cheese,sauce,slice,thin,like,wing,great,topping \\ \cline{2-3} 
 & Food (Salad) & salad,bread,cheese,garlic,tomato,fresh,sauce,olive,delicious,oil \\ \cline{2-3} 
 & Service & service,staff,friendly,server,owner,customer,waiter,always,attentive \\ \cline{2-3} 
 & Location & place,restaurant,location,strip,little,find,italian,away,parking,right \\ \cline{2-3} 
 & Value & food,good,price,better,much,pretty,like,quality,portion,worth,nothing \\ \hline
\multirow{5}{*}{French} & Drinks & menu,wine,course,tasting,glass,bottle,ordered,selection,meal,two \\ \cline{2-3} 
 & Dessert & dessert,chocolate,cream,cake,ice,coffee,sweet,creme,tart,also,good,souffle \\ \cline{2-3} 
 & Food (Bread) & bread,egg,french,butter,good,toast,delicious,fry,cheese,fresh,croque \\ \cline{2-3} 
 & Food & cheese,salad,soup,onion,ordered,good,french,delicious,appetizer,lobster \\ \cline{2-3} 
 & Service Time & table,minute,time,wait,reservation,waiter,get,seated,server,took,order,got \\ \hline
\end{tabular}
\caption{Local topics for Indian, Italian and French restaurants obtained using MG-LDA}
\label{table:mglda}
\end{table*}
\FloatBarrier

\noindent modeling methods such as LDA and PLSA, which extract topics that correspond to global properties of a product, MG-LDA extracts much finer topics that correspond to ratable aspects of the product. To extract topics at such granular level, the model generates terms which are either chosen at the document level or chosen from a sliding window\footnote{A sliding window is a set of fixed number of adjacent sentences.}. The terms chosen from the sliding window correspond to the fine topics.

\subsection{Review Segmentation and Sentiment Analysis}
\label{ssec:mapping}

Once cuisine-specific latent topics are obtained, the review segmentation task is performed where each review sentence $s_i$ is assigned to one of the latent topics $t_k$. The purpose of this task is to understand which sentences of the review discuss which of the topics. The topic assignment is made as follows:
\begin{equation}
 Topic(s_i)=\operatorname*{arg\,max}_k \sum_{w\in{t_k}}count (w, s_i)  *  P(w|t_k)
\end{equation}
where w is the word associated with each topic, count(w, $s_i$) is the count of word w in sentence $s_i$ and $P(w|t_k)$ is the probability as determined from the word distributions obtained using the MG-LDA model.

For every review, the sentences that discuss each topic are identified as shown in Figure~\ref{figure:segmentation}. It is therefore possible to determine the sentiment expressed by the review author regarding each latent topic by averaging over the sentiments of its constituent sentences. We use the implementation TextBlob\footnote{\url{http://www.textblob.readthedocs.org/en/dev/}}, which is based on the Pattern\footnote{\url{http://www.clips.ua.ac.be/pattern}} library, to determine the polarity of each sentence. The polarity is obtained in the range of [-1, 1]. 

\subsection{User Segmentation}
\label{ssec:rep. users}

We then proceed to learn the representative users for each latent topic. First, the feature vector $\theta^{overall}_u$ is obtained for each user $u$ where each feature represents the users' review rating for a restaurant. We assume that each user writes only one review per restaurant. Similarly, $\theta^{t_k}_u$ is obtained where each feature represents the users' sentiment regarding topic $t_k$. 

The influence of a topic on a users' rating is determined by calculating the Pearson's correlation between $\theta^{overall}_u$ and $\theta^{t_k}_u$. Only users who have provided a minimum of 5 reviews are considered. A user-topic correlation matrix $C$ is thus obtained which indicates the dimensions along which each

\FloatBarrier
\begin{figure*}[t]
\centering
\includegraphics[scale=0.5]{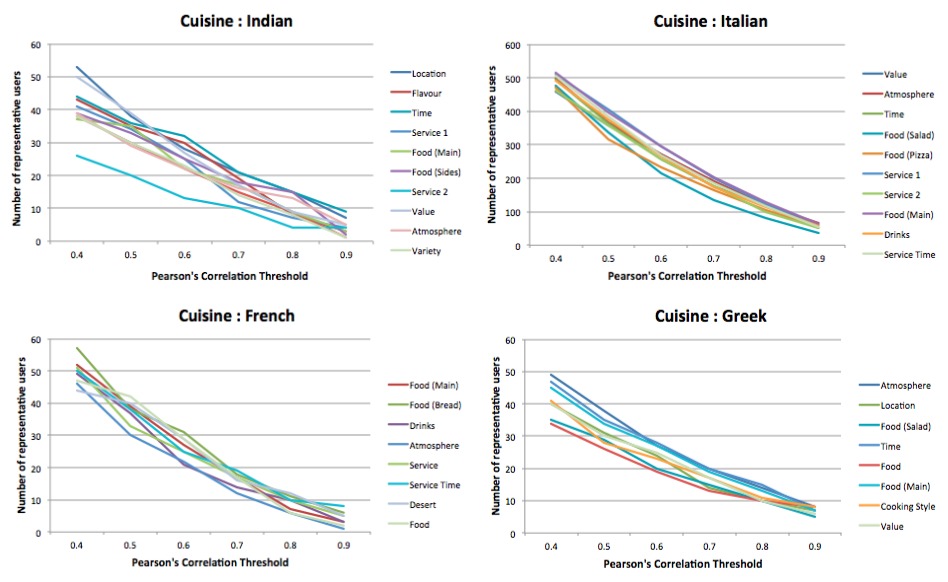}
\caption{Number of Representative Users for various cuisines}
\label{figure:graphs}
\end{figure*}
\FloatBarrier

\noindent user tends to rate restaurants. Simply put,
\begin{equation}
 C(u,t_k)=PearsonCorr(\theta^{overall}_u,\theta^{t_k}_u)
\end{equation}
The representative users for a topic are those users whose $C(u,t_k)$ value is above a certain threshold $T$ for that particular topic. It is important to note that $C(u,t_k)$ value may not be available for all user-topic pairs since every user may not express sentiments regarding every topic. 

\subsection{Rating Prediction}
\label{ssec:prediction}

We calculate the topic ratings of restaurants once we obtain a list of representative users for each latent topic. This rating is calculated as the average of the review ratings that are provided by the representative users of that particular topic. In case there are no representative users for a particular topic for that particular restaurant, this rating is calculated as the average of the other latent topic ratings. Such topic ratings provide some indication of the quality of various aspects of the restaurant (like food, service etc.), although we do not explicitly calculate the aspect ratings or map the topics to aspects. 

Since the overall restaurant rating can be thought of as some combination of the ratings for food, service, atmosphere etc., we try to combine the latent topic ratings in some way. For this purpose, we fit a Support Vector Regression (SVR) model with radial basis function kernel on the latent topic ratings and use it to predict the overall rating of restaurants. During test time, just the ratings provided by a few representative users would be enough to obtain the overall restaurant rating. Such a rating takes into account the different dimensions of the restaurant and provides a `complete picture' of the restaurant. 

\section{Experiments and Analysis}
\label{sec:experiments}

We use the topic model MG-LDA on a set of 8000 reviews each of Indian, Italian, French and Greek restaurants. The number of global topics is set at $K_{glo}=40$ and local topics at $K_{loc}=15$ (After trying various combinations, we found that this combination provides the best results. Previous works have also used a similar number of topics). The length of the sliding window is set at 2 and all the other parameters for the model is set at 0.1. We run the chain for 1000 iterations. While the global topics are ignored, some select local topics as determined by the model are shown in Table~\ref{table:mglda}. We try to interpret the topics manually by looking at the constituent words. Usually, around 5-6 local topics are ambiguous and difficult to interpret. 

A quick look at the topics obtained shows us the variation that exists among different cuisines. For example: While Indian restaurants have `Flavour' and `Variety' as topics; Italian restaurants have `Drinks'; French restaurants have `Drinks' and `Dessert' as topics. Greek restaurants have `Cooking Style' as a topic with words like dry, fry, fresh, cooked, soft, tender etc. Also, certain words like table, minute, time, wait, hour, bar, seated etc. appear together in case of French and Italian restaurants signaling, perhaps, a long wait to get seated at such restaurants.

Review segmentation is then performed on around 8500 reviews of Indian restaurants, 61000 reviews of Italian restaurants and 17000 reviews of French restaurants, where each sentence is assigned to one of the 15 latent topics. Sentiment analysis is conducted and the user-topic correlation matrix is obtained for each restaurant category. 

Using the user-topic correlation matrix, we segment the users according to each latent topic. Figure~\ref{figure:graphs} shows the number of representative users for each topic for different correlation thresholds $T$. For the sake of clarity, we only show those latent topics that could be interpreted by us. It is interesting to observe that people who visit Indian restaurants tend to care the most about `Location' and `Value (Pricing)' and the least about `Service' and `Atmosphere'. On the other hand, people who visit French restaurants care the most about `Food (Bread)' and `Food (Main)' and the least about `Atmosphere'. Similarly, while providing ratings, more number of users are influenced by the `Atmosphere' at Greek restaurants than `Food'. We then proceed to obtain the latent topic ratings for each restaurant. For this purpose, we only select those users whose correlation threshold, $T >= 0.4$ as representative users. For each latent topic, we average over the ratings provided by such users to obtain the topic ratings (out of 5). It is therefore possible to obtain crude ratings for aspects like `Food', `Service' etc. which give an indication of the quality of the aspects. We then fit an SVR model, the performance of which is described below.

\subsection{Evaluation}
\label{ssec:eval}

To evaluate the performance of rating prediction, we determine the RMSE between the actual and predicted ratings for Italian restaurants. We compare the RMSE for MG-LDA and online LDA described in \cite{hoffman2010online}. In case of LDA, we detect $K = 50$ topics as in previous works. We use the latent topic ratings of 640 restaurants for training and 215 restaurants for test. The results are shown in Table~\ref{table:eval1}.

\FloatBarrier
\begin{table}[h]
\begin{center}
\begin{tabular}{|l|c|}
\hline
\multicolumn{1}{|c|}{Models}                                                                  & RMSE   \\ \hline
\begin{tabular}[c]{@{}l@{}}(a) MG-LDA, SVR with rbf kernel\\ (Proposed Model)\end{tabular} & 0.4909 \\ \hline
(b) MG-LDA, SVR with linear kernel                                                            & 0.5377 \\ \hline
(c) LDA, SVR with rbf kernel                                                                  & 0.5812 \\ \hline
(d) LDA, SVR with linear kernel                                                               & 0.6277 \\ \hline
(e) Baseline 1                                                                                & 0.6737 \\ \hline
(f) Baseline 2                                                                                & 0.5831 \\ \hline
\multicolumn{2}{|c|}{Improvement}                                                                      \\ \hline
\multicolumn{1}{|c|}{(a) vs. (e)}                                                             & 27\%   \\ \hline
\multicolumn{1}{|c|}{(a) vs. (f)}                                                             & 16\%   \\ \hline
\end{tabular}
\end{center}
\caption{Evaluation (Italian Restaurants)}
\label{table:eval1}
\end{table}
\FloatBarrier

An RMSE of 0.4909 is obtained when using MG-LDA and SVR with rbf kernel. Each restaurant has an average of 22 representative users. Inviting these users to rate new restaurants would help in predicting the `true' restaurant rating (which is the rating obtained once a considerable number of users have rated the restaurant over a period of time). However, conventional methods just average over their ratings, without taking into account the different topics that they represent. Such an approach gives an RMSE of 0.6737 (Baseline 1). Our approach outperforms this method by 27\%. Also, since most people provide a rating of 3, 3.5 or 4 when rating restaurants, predicting a constant rating every time may also give a reasonable result. We find that predicting a rating of 3.64 (average over the test set) every time results in an RMSE of 0.5831 (Baseline 2). Our approach outperforms such a constant classifier by 16\%.

We repeat the same procedure for Indian restaurants by using the latent topic ratings of 120 restaurants for training and 40 restaurants for test. The results are shown in Table~\ref{table:eval2}.

\FloatBarrier
\begin{table}[h]
\begin{center}
\begin{tabular}{|c|c|}
\hline
Models                         & RMSE   \\ \hline
MG-LDA, SVR with rbf kernel    & 0.4635 \\ \hline
MG-LDA, SVR with linear kernel & 0.5795 \\ \hline
LDA, SVR with rbf kernel       & 0.5734 \\ \hline
LDA, SVR with linear kernel    & 0.6997 \\ \hline
\end{tabular}
\end{center}
\caption{Evaluation (Indian Restaurants)}
\label{table:eval2}
\end{table}
\FloatBarrier

\section{Conclusion and Future Work}
\label{sec:end}

In summary, we show how latent topics in review text could be used to unlock hidden value in user reviews. We utilise the intuition that, while rating products, certain users are influenced heavily by one particular aspect of the product. We learn such users by detecting the sentiments expressed by them with regard to each latent topic and then by comparing these sentiments with the actual ratings provided. We also use this to draw some interesting insights regarding users' rating behaviour across different cuisines and obtain latent topic ratings for restaurants. Overall ratings, which take into account the different dimensions of the restaurant, are then obtained using a regression model. 

In the future, we would like to show that this approach is transferable to other domains like e-commerce. Also, it would be interesting to segregate the reviews by star ratings as this would help us understand the factors that a restaurant is getting right and those they are getting wrong. For example: The dimensions corresponding to review text having 5-star ratings would be different from those having 1-star ratings.

\bibliographystyle{acl}
\bibliography{bib}

\end{document}